# A Pathway for Assessing Grey Literature: Leveraging AI to Extract Conference Metadata and Organiser Information from Calls for Papers

Angelo Salatino*, Francesco Osborne*, Alexis Vizcaino**, Aliaksandr Birukou**, and Enrico Motta*

* *angelo.salatino@open.ac.uk, francesco.osborne@open.ac.uk, enrico.motta@open.ac.uk*
0000-0002-4763-3943; 0000-0001-6557-3131; 0000-0003-0015-1952
Knowledge Media Institute, The Open University, UK

** *alexis.vizcaino@springernature.com; aliaksandr.birukou@springer.com*
0000-0003-4364-284X; 0000-0002-4925-9131
Springer Nature, Germany

Despite its importance, grey literature, including Calls for Papers (CfPs), remains largely overlooked in Metascience and Scientometric analysis due to its unstructured, highly heterogeneous format, which traditional tools struggle to process at scale. However, Large Language Models now offer a pivotal opportunity to devise innovative tools for systematically harvesting and processing such data. In this paper, we introduce COCI, an AI-based framework that automates the extraction of granular, structured metadata from raw CfP text. COCI employs a multi-stage pipeline for entity extraction, followed by author disambiguation against OpenAlex and semantic mapping of topics and conference series. This process identifies key data points, including conference editions, geographic locations, and comprehensive lists of organisers, along with their specific roles and affiliations. By structuring this previously inaccessible information, COCI establishes a foundation for the systematic analysis of grey literature, enabling new research opportunities and shifting the scholarly focus towards non-publisher-based events.

## 1. Introduction

The majority of Metascience and Scientometric analyses focus on extracting insights from the published literature, typically available through journal issues or conference proceedings. Such studies have yielded significant understanding of how science evolves (Klavans and Boyack, 2006), how research teams are composed (Salatino et al., 2026), and how to detect emerging trends (Salatino et al., 2018), while also enabling practical solutions like recommenders (Thanapalasingam et al., 2018) and automated literature reviews (Bolaños et al., 2024). However, a vital dimension of research, namely grey literature, remains largely overlooked (Kousha et al., 2022). Produced by organisations outside traditional commercial or academic publishing channels, this category covers reports, working papers, government documents, white papers, and theses (Adams et al., 2017). Crucially, Calls for Papers (CFPs) also perfectly embody the defining characteristics of grey literature, as they are knowledge artefacts that lack traditional bibliographic control, bypass the standard peer-review process, and are frequently distributed through informal channels, including personal websites, emails, and community newsletters. While such literature is essential for accessing up-to-date, specialised information, it is often scattered across the web in highly heterogeneous formats (Schöpfel and Farace, 2009). This lack of structure makes it increasingly difficult to analyse them at scale.

The emergence of Large Language Models (LLMs) represents a significant turning point. We have now reached a stage that enables more ambitious scientometric visions, allowing the development of innovative tools to address the technological limitations that traditional AI methods encountered when harvesting and processing grey literature.

In line with this objective, we present the Conference Organisers and Content Identifier (COCI), an AI-based framework designed to automate the extraction of fine-grained metadata from unstructured CfPs. COCI processes raw CfP text to identify and structure key information, including conference series and specific editions, acronyms, geographic locations, and comprehensive lists of organisers. In addition, the system identifies organiser affiliations and roles, such as research, workshop, and publicity chairs, as well as the thematic topics of interest. The extracted data is structured into a standardised format and presented via an intuitive user interface.
In summary, COCI facilitates the analysis of grey literature by extracting rich metadata, including conference URLs that serve as entry points for further discovery. This foundation enables the deployment of AI agents capable of crawling event websites to capture diverse research artefacts, such as preprints, presentations, and videos. By automating processes that were previously limited by technical barriers, COCI provides a pathway toward a more inclusive and comprehensive understanding of the research landscape. To support reproducibility and community-driven research, COCI is released as an open-source project[1].

## 2. Related Work

We review the literature concerning the core challenges and techniques for extracting conference organisers and identifying scholarly content.

### *2.1. Information Extraction from Scholarly Text*

Information extraction (IE) aims to derive structured representations, such as entities, attributes, and relations, from unstructured or semi-structured text, and it underpins services like digital libraries, knowledge graphs, and analytics on the evolution of scientific communities (Salatino et al., 2019a, 2019b; Turki et al., 2021). Beyond full articles, artefacts such as proceedings titles and CfPs encode rich event metadata but lack consistent structure (Franken et al., 2022; Lackner et al., 2021). Traditional scholarly IE systems relied on rule-based heuristics, regular expressions, and classical machine learning to extract bibliographic and event-related fields, which work well with stable templates but degrade with heterogeneous or noisy sources, such as publisher-specific layouts or *ad hoc* CfP and web pages (Hong et al., 2021). Indeed, Fahl et al. (2023) demonstrate that conference metadata, such as series title, year, and location, can be automatically extracted from thousands of CEUR volumes. However, without cooperation between researchers and publishers, such an approach cannot automatically adapt to new document genres, event types, and community conventions (Bryl et al., 2014).
LLMs transformed scholarly IE by reframing extraction as conditional text generation. In the scientific domains, LLMs extract study characteristics and knowledge with minimal task-specific supervision, provided that prompts specify target fields and include in-context examples (Li et al., 2025). At the same time, LLM-based extraction remains highly sensitive to prompt wording and is vulnerable to hallucinations. In the context of our work, the LLM must interpret informal phrasing, infer implicit information such as event series names, and adapt quickly to new schemas.

### *2.2. Author and Entity Disambiguation*

Author and entity disambiguation represents a foundational challenge in bibliometric research, as the accuracy of analyses and indicators depends on correctly identifying and linking authors and institutions across datasets (Cappelli et al., 2025). This task has long relied on heuristic and machine learning approaches to distinguish between individuals with identical or similar names (Santini et al., 2022). Traditional methods used rule-based systems and metadata similarity based on co-authorship networks, publication venues, and topical analysis (D'Angelo et al., 2011; Tekles and Bornmann, 2020), while recently machine learning efforts leveraged embeddings and graph-based learning (Bonis et al., 2023; Rastogi et al., 2023). However, despite these advances, errors remain common, particularly for early-career researchers, authors with frequent affiliation changes, or those working in multidisciplinary fields. Persistent identifiers like ORCID and infrastructures such as OpenAlex improve linking reliability, though these systems still depend on consistent metadata curation (Auhunas et al., 2024; OpenAlex, 2023).

[1] COCI GitHub repository — https://github.com/angelosalatino/oc-conf-detection

Affiliation disambiguation poses similar difficulties because institutions often appear under multiple names, abbreviations, or different languages (Chen et al., 2023). Resources like the Research Organization Registry[2] (ROR) provide standardised, open identifiers. Matching algorithms in this domain typically combine string similarity with geographic information to align affiliation strings with ROR records (Tkaczyk and Gould, 2019). Nevertheless, affiliation disambiguation remains a challenging task, especially for multi-campus universities, industrial partners, or research consortia that evolve over time (Purnell, 2022). Quality assurance now includes cross-validation with trusted sources and automated anomaly detection to identify misattributed authors or non-existent entities (Purnell, 2022).

### *2.3. Controlled Vocabulary Mapping*

Mapping free-text terms to structured conceptual frameworks is essential for semantic interoperability (Salatino et al., 2025). Traditional approaches rely heavily on lexical similarity, where string-matching techniques such as Levenshtein distance or term co-occurrence statistics are used to align user-provided topics with entries in controlled vocabularies like the OpenAlex Concepts (Ampudia Vicente, 2025) or the Computer Science Ontology (Osborne et al., 2016; Salatino et al., 2019b). While adequate for exact matches, these methods struggle with paraphrasing or domain-specific terminology.

Recent studies employ semantic mapping grounded in language models. Embedding-based methods, such as those using SentenceTransformers (Reimers and Gurevych, 2019), represent terms as dense vectors, allowing similarity to be computed through cosine similarity. This enables the discovery of concept alignments beyond surface lexical cues. For instance, the topic “machine perception” may be automatically associated with “computer vision” due to their proximity in the embedding space. This approach bridges gaps between informal representations and formal taxonomies. Hybrid strategies combining lexical matching and semantic similarity are increasingly common, balancing precision with the robustness of contextual embeddings (Salatino et al., 2019b).

### *2.4. Conference Series and Event Alignment*

A growing body of work in bibliometrics has focused on aligning conference series across data sources to track longitudinal research trends (Angioni et al., 2021). Major infrastructures provide structured metadata for scientific conferences, including DBLP, which offers a long‑standing, curated registry of computer‑science proceedings and conference series, and newer platforms like ConfIDent (Hagemann-Wilholt et al., 2020), which aim to model events as persistent, citable entities linked to broader scholarly knowledge graphs. In this space, the AIDA Dashboard integrates publication and event information to enable comparative analytics of the research landscape (Angioni et al., 2022).

Aligning conferences across diverse datasets remains a key challenge (Zhang et al., 2025). Syntactic methods matching surface features, such as acronyms, ensure high precision but suffer from low recall when naming variations are complex (Schraagen, 2014). To address this, semantic approaches utilise formal ontologies and LLMs to identify shared research themes and synonymous entities (Thießen et al., 2023; Vahdati et al., 2015). Ultimately, hybrid strategies combining both syntactic and semantic dimensions achieve robust, highly precise alignment (Salatino et al., 2019b). Our proposed approach adopts this hybrid paradigm, utilising SentenceTransformers (Reimers and Gurevych, 2019) to broadly identify close matches with high recall, before applying syntactic similarity to ensure high precision.

[2] Research Organization Registry — https://ror.org/

### *2.5. Analysing Grey Literature*

Scholarly communication has expanded beyond traditional proceedings to include grey literature such as preprints, technical reports, and CfPs (Paez, 2017; Osayande and Ukpebor, 2012). These artefacts capture emerging research directions before they reach formal publication channels. However, the informal nature of such material introduces challenges for systematic tracking. Unlike peer-reviewed papers, grey literature is seldom indexed in standardised databases, and its metadata may be scattered across institutional repositories (Paez, 2017).

This project reorients the analytical lens toward non-publisher-based events as pivotal fora of exchange. Conferences lacking formal proceedings, including STI-ENID[3], provide insights into research trends otherwise invisible in traditional analyses. By positioning CfPs as a key source of evidence, this work acknowledges their role in revealing emerging directions and novel paradigms within academic communities. Analysing CfPs enables the identification of disciplinary intersections and collaboration networks at their inception, offering a bottom-up perspective on scholarly communication that embraces the diversity of modern research dissemination.

## 3. Approach

COCI processes unstructured *Calls for Papers* text to extract structured metadata, identify organisers, and link entities to external services. The workflow, reported in Figure 1, consists of five main stages: i) prompt engineering for LLMs extraction, ii) author disambiguation against OpenAlex, iii) conference series matching, iv) topic mapping, and v) data visualisation. While LLMs facilitate the initial extraction, their raw outputs often lack the precision required for scholarly databases. Therefore, COCI is designed as a multi-stage methodology rather than a simple prompt interface. It treats LLM data as a starting point, employing rigorous post-processing and external validation to ensure the resulting metadata is robust, disambiguated, and reliable.

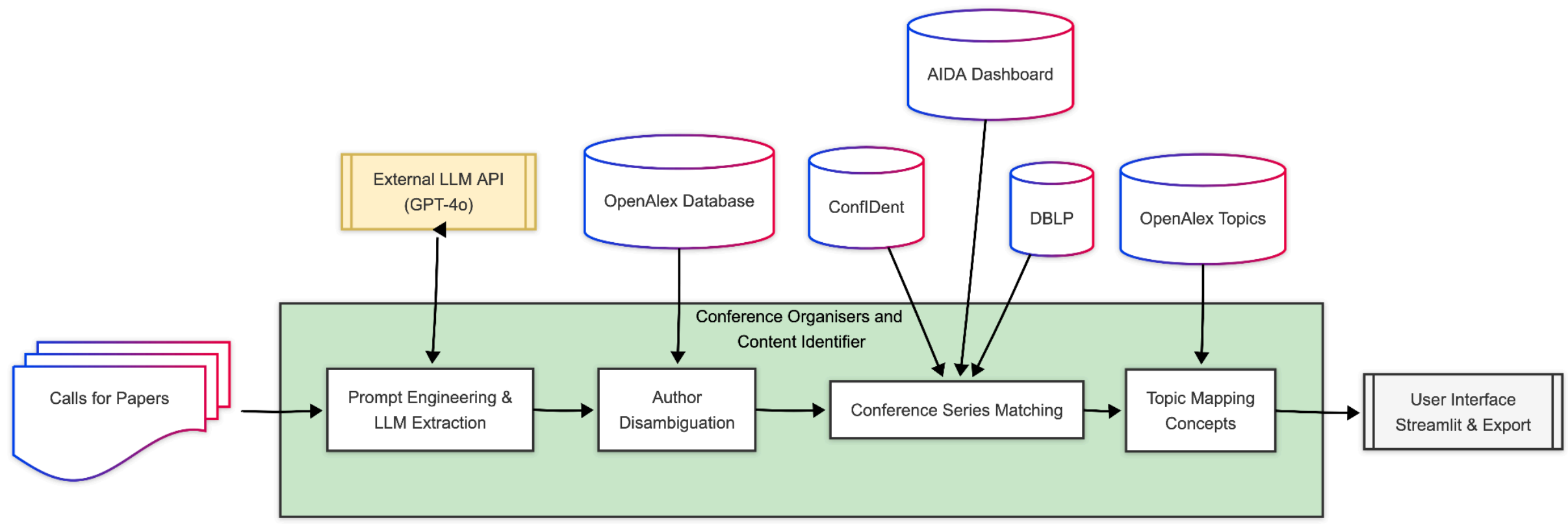

Figure 1: Main workflow of the Conference Organisers and Content Identifier.

[3] It is important to note that while STI-ENID publishes formal proceedings, their location varies by edition, appearing across platforms such as Zenodo, Orvium, or the official ENID-Europe website, which makes the harvesting process significantly more challenging. Source https://enid-europe.eu/index.php?id=confProceedings

### *3.1. Prompt Generation*

To extract structured information from the raw text of a CfP, we employ the GPT-4o model developed by OpenAI. We utilise a prompt template refined over several iterations to handle the variability in CfPs across different scientific disciplines.

The prompt instructs the model to parse the input text and extract specific entities, including the event name, acronym, conference series, year, location, and a comprehensive list of topics of interest. A critical component of the prompt is the exhaustive extraction of the organising committee. The model is directed to identify every person listed, capturing their full name, affiliation, country, and specific track or role (e.g., "General Chair", "Research Track").

To ensure the output is machine-readable and strictly typed, we enforce a JSON schema in the API request. This schema defines the expected structure for nested objects, such as the list of organisers, and ensures that fields like year or topics adhere to specific formats.

### *3.2. Author Matching*

Once the raw list of organisers is extracted, the system attempts to link each individual to an author profile in the OpenAlex database. This process enriches the data with persistent identifiers, such as ORCIDs, OpenAlex IDs, and Research Organization Registry (ROR) IDs for affiliations. The matching process follows a multi-stage strategy.

The search strategy prioritises precision by first attempting to locate the organiser's institution within OpenAlex, then filtering for authors linked to that specific institutional ID. During this stage, COCI also assesses whether the extracted affiliations are suspiciously homogeneous, suggesting a potential LLM hallucination in which the model has repeated the same institution for every member. In this case, the affiliation data is discarded to maintain integrity.

Should the affiliation-based search fail, or if no affiliation was provided, the system reverts to a name-only search. To disambiguate common names, candidates are ranked using heuristics that favour researchers with a high number of works. This involves analysing publication counts and employing Levenshtein string similarity against known name variations to identify the most likely profile.

Following identification, the system performs a final verification and inference step. When an affiliation is present in the CfP, it is fuzzy matched against the organiser's publication history to confirm accuracy and retrieve the relevant ROR ID. If this matches their OpenAlex profile, the record receives a *verified* badge in the user interface, akin to social media verification. In the absence of a provided affiliation, the system infers the most likely current institution by analysing the author's publications over the past decade, prioritising academic institutions over other types of affiliations, such as commercial or governmental bodies.

### *3.3. Topics of Interest*

The raw topics of interest extracted by the LLM are often unstructured or non-standardised. To facilitate better indexing and analysis, we map these terms to the controlled vocabulary of OpenAlex concepts.

To achieve this, all extracted topics are converted into 384-dimensional dense vector embeddings using the "all-MiniLM-L6-v2"[4] SentenceTransformers model. These embeddings map semantically similar concepts to proximal locations within a multidimensional space, ensuring that semantically distant concepts remain distinct. We used cosine similarity to assess the proximity between entities, applying a similarity threshold to identify semantically equivalent terms. In the current version, this threshold is set to 0.6.

The topic embeddings are then compared against an index of pre-computed OpenAlex concept embeddings. This methodology enables the seamless integration of topics of interest with the OpenAlex taxonomy while avoiding the pitfalls of rigid syntactic comparison. By moving beyond literal string matching, the system ensures that minor variations, such as hyphenation or pluralisation, do not result in a mismatch between identical concepts. Ultimately, this pipeline transforms a list of ad-hoc keywords into a set of standardised and well-defined research concepts.

[4]Sentence Transformers — https://huggingface.co/sentence-transformers/all-MiniLM-L6-v2

### *3.4. Conference Series Matching*

To accurately situate each event within the broader academic landscape, the extracted conference series names are matched against three primary external services: DBLP[5], the AIDA Dashboard[6], and ConfIDent[7]. This alignment is performed through semantic search using sentence transformers, following the same approach adopted for identifying the topics of interest. The resulting embeddings are queried against precomputed indices for each service, enabling the system to retrieve the nearest neighbours based on semantic similarity.
To ensure high precision, the semantic matches are validated using Levenshtein string similarity. Furthermore, it leverages cross-referencing capabilities. For instance, if a match is found in DBLP, the system checks internal mappings to automatically retrieve the corresponding identifiers for AIDA and ConfIDent, ensuring a cohesive set of external links. The dual-layer approach of semantic embeddings followed by Levenshtein validation effectively mitigates the risks of false positives often found in purely vector-based retrieval.

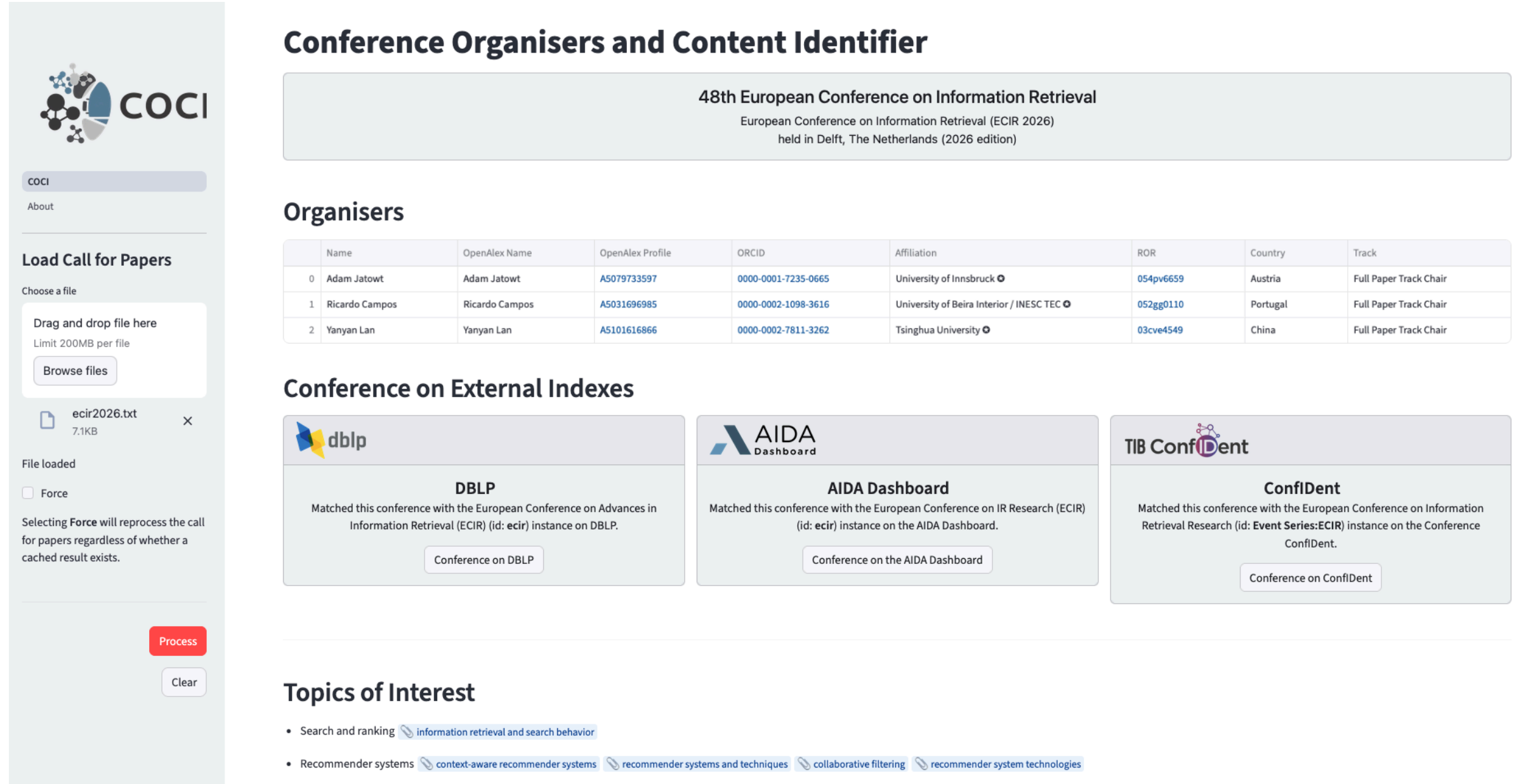

Figure 2: Graphical user interface of COCI.

### *3.5. Visualisation*

The processed data is presented via a web interface built using Streamlit[8] (Figure 2). The sidebar allows users to load the CfP, whereas the main panel provides a structured overview of the conference, including its metadata, organisers, and external records.
Specifically, the main panel first details the conference series, edition, location, year, and acronym. Then it displays an interactive table of the organising committee. This table maps extracted names to their matched OpenAlex profiles, displaying specific roles, ORCIDs, and ROR-linked affiliations. The ✪ symbol indicates that the affiliation found in the CfP aligns with the OpenAlex author profile, serving as a verified measure of confidence in the match. To further contextualise the event, the system displays links to external records from DBLP, AIDA, and ConfIDent, alongside thematic topics mapped to the OpenAlex taxonomy (highlighted in blue). Finally, users can export the entire enriched dataset as a Microsoft Excel file for offline analysis.

---

[5] DBLP — https://dblp.org/
[6] AIDA Dashboard — https://aida.kmi.open.ac.uk/dashboard
[7] TIB ConfIDent — https://www.confident-conference.org
[8] Streamlit — https://streamlit.io/

## 4. Evaluation

We evaluated COCI by analysing the quality of the extracted metadata. We defer the evaluation of the user interface, as the current version is primarily intended to support functional code execution and is likely to undergo a complete redesign. Accordingly, our analysis focuses on the technical performance of COCI.

Specifically, we processed 40 CfPs and manually inspected the outputs. The CfPs were drawn from a diverse range of fields, including Computer Science, Engineering, Materials Science, Information Science, Earth and Planetary Science, Dentistry and Oral Health, and Archaeology and History. This disciplinary breadth allowed us to assess whether COCI generalises effectively across different academic disciplines. This evaluation followed an iterative test and learn approach, whereby we refined our prompt engineering and validation logic in response to initial results, and reached a stage of functional maturity where COCI successfully processed all CfPs and extracted their metadata.

However, through the evaluation we identified specific inconsistencies that highlight the inherent challenges of this task, which we report as representative examples.

The first occurred when processing the call for papers for STI 2026, where the source material did not provide affiliations for the organising committee. In this instance, the LLM hallucinated, incorrectly attributing all members to the University of Antwerp (the local organiser). Given that conferences where the entire committee originates from a single institution are exceedingly rare, we introduced a heuristic to evaluate affiliation distribution. This ensures diversity by verifying that the number of unique organisers does not exceed four times the number of unique organisations, otherwise the affiliation data is deemed unreliable.

In another instance, while processing the 24th International Semantic Web Conference, one organiser, Andrea Giovanni Nuzzolese, was incorrectly matched with an author named Isabel F. Cruz in OpenAlex (ID A5104780631[9]). Further investigation revealed that this error stems from the OpenAlex database itself, which lists “Andrea Giovanni Nuzzolese” as an alternative name for Dr Cruz. This suggests that certain inaccuracies are contingent upon the quality of external data rather than a failure of the matching algorithm itself. Finally, we noted instances where the LLM hallucinated organisers’ roles or failed to comprehend specific roles defined in the text, defaulting instead to “main track”. Following further prompt engineering, this behaviour has been stabilised to ensure more accurate role identification.

## 5. Vision and Future Plan

As previously noted, this project was conceived to analyse conferences beyond those that exclusively publish formal proceedings, shifting the focus towards grey literature and alternative scholarly outputs. This endeavour presents a distinct challenge: conferences operating outside the traditional proceedings-based model lack both centralised publishers and standardised bibliographic records. Consequently, gathering reliable data for these events introduces a significantly higher level of complexity compared to traditional, indexed conferences. For instance, while some events solicit abstracts through open calls, others operate strictly on an invitation-only basis. Furthermore, outputs vary significantly, with some conferences publishing video recordings (e.g., VideoLectures.net) or depositing alternative materials into online repositories (e.g., Zenodo Communities).
Given this highly heterogeneous landscape, the analysis of calls for papers constitutes a natural starting point. These documents contain a rich body of information that, once harvested and processed, can support a wide range of research and analytical applications. Ideally, CfPs serve as the primary gateway for deeper exploration. By extracting the conference URL from the call, an automated AI agent can navigate the event's website to identify and capture additional grey literature worthy of analysis.
Beyond merely parsing and structuring basic conference data, COCI can become a platform to assess the health and scientific quality of these events, evaluating whether a conference is organised by a diverse and appropriately senior team, and whether it maintains a well-defined scope. Furthermore, connecting it to external databases such as RetractionWatchDatabase and PubPeer allows users to investigate whether organisers have previously engaged in misconduct.
Finally, while exhaustive web crawling is impractical, in our vision, COCI will remain scalable by tracking established CfP registries (e.g., WikiCFP) and professional mailing lists (e.g., ISSI Society), supplemented by a community-driven submission system to steadily expand its coverage. Furthermore, by mapping the professional backgrounds of organisers and their relationships to journals, editorial boards, and proceedings editors, COCI provides the infrastructure for a recognition platform. Ultimately, we envision this dataset enabling a system that grants researchers formal credit for their community contributions, analogous to the model established by Publons for peer review.

---

[9] Isabel F. Cruz on OpenAlex — https://web.archive.org/web/20260213174312/https://openalex.org/authors/A5104780631 (page has been web archived to ensure long-term preservation and to provide a verified point-in-time reference).

## 6. Conclusions

In this paper, we introduced COCI, a novel AI-based framework designed to systematically analyse CfPs. This tool, based on recent developments in AI, can effectively extract granular, structured metadata from highly heterogeneous CfP text. COCI's core contributions stem from its ability to automate entity extraction, perform author disambiguation by linking to external knowledge bases such as OpenAlex, and semantically map topics and conference series names to achieve standardisation. By transforming unstructured data into a rich and structured dataset, COCI enables a new generation of scientometric analyses focused on scholarly events beyond traditional publisher-based venues.
Looking ahead, we plan to significantly extend COCI's functionalities to fully realise its potential as a systematic analysis platform for the grey literature landscape. Our immediate future work will focus on developing an automatic, continuous crawling engine capable of harvesting new calls for papers from a diverse set of registries and mailing lists, ensuring a constantly updated and comprehensive data source. Following this, we intend to design and implement a systematic method to organise all parsed events, enabling longitudinal and comparative analysis of conference series, topics, and communities. Finally, we plan to develop new platforms and downstream applications that leverage the harvested and enriched data, unlocking novel research and analytical use cases for the scholarly community.

## Open science practices

In line with our commitment to open science, the COCI framework is released as an open-source project. The software and its source code, as well as the data (the 40 Calls for Papers) used for the evaluation presented in this paper, are publicly available on our GitHub repository. To ensure the reproducibility of our technical performance claims, we have created the v1.0.0 release. By downloading this specific release, any researcher can access the exact version of the code and data used in our experiments, allowing them to reproduce our reported outputs and results. The repository, including the preservation release, can be accessed here: https://github.com/angelosalatino/oc-conf-detection/releases/tag/v1.0.0.

**Author contributions**

The authors contributed to the work as follows, according to the CRediT system: **Conceptualization**: Angelo Salatino; **Methodology**: Angelo Salatino; **Software**: Angelo Salatino; **Validation**: Angelo Salatino; **Funding acquisition**: Alexis Vizcaino, Aliaksandr Birukou, **Writing – Original Draft**: Angelo Salatino; **Writing – Review & Editing**: Angelo Salatino, Francesco Osborne, Alexis Vizcaino, Aliaksandr Birukou, and Enrico Motta.

**Competing interests**

The authors declare that they have no competing interests.

**Funding information**

We would like to thank Springer Nature for funding this research.